\documentstyle[psfig,floats,twocolumn,aps,prl]{revtex}

\def\w{\omega}

\newcommand{\R}{\ifmmode{I\hskip -4pt R}
\else{\hbox{$I\hskip -4pt R$}}\fi}
\newcommand{\N}{\ifmmode{I\hskip -4pt N}
\else{\hbox{$I\hskip -4pt N$}}\fi}

\def\div{{\rm div}}

\begin{document}
\ifx\undefined\psfig\def\psfig#1{ }\else\fi
\ifpreprintsty\else
\twocolumn[\hsize\textwidth%
\columnwidth\hsize\csname@twocolumnfalse\endcsname
\fi
\title{Two-Dimensional Modeling of Soft Ferromagnetic Films}
\author{Antonio DeSimone,$^{1}$ Robert V. Kohn,$^2$ Stefan M\"{u}ller,$^1$ 
Felix Otto,$^3$ and Rudolf Sch\"{a}fer$^4$}
\address{$^1$Max-Planck-Institute for Mathematics in the Sciences, Inselstr. 22-26,
D-04103 Leipzig, Germany}
\address{$^2$Courant Institute of Mathematical Sciences, 251 Mercer St.,
New York, NY 10012}
\address{$^3$Institut f\"{u}r Angewandte Mathematik, Universit\"{a}t Bonn, Wegelerstr. 10,
D-53115 Bonn, Germany}
\address{$^4$Institut f\"{u}r Festk\"{o}rper- und Werkstofforschung,
Helmholtzstr. 20,
D-01069 Dresden, Germany}
\date{Leipzig, April 30, 2000}
\maketitle
\begin{abstract}
We examine the response of a soft ferromagnetic film to an in-plane applied
magnetic field. Our theory, based on asymptotic analysis of the micromagnetic
energy in the thin-film limit, proceeds in two steps: first we determine the
magnetic charge density by solving a convex variational problem; then we
construct an associated magnetization field using a robust numerical method.
Experimental results show good agreement with the theory. Our analysis is
consistent with prior work by van den Berg and by Bryant and Suhl, 
but it goes much further; in
particular it applies even for large fields which penetrate the sample.
\end{abstract}
\maketitle
\ifpreprintsty\clearpage\else\vskip1pc]\fi
\narrowtext
Soft ferromagnetic films are of great interest both for applications
and as a model physical system. Their sensitive response to applied
magnetic fields makes them useful for the design of many devices,
including sensors and magnetoelectronic memory elements \cite{applications}.
Therefore soft thin films have
been the object of much experimental and computational study
\cite{HS}. Their relatively simple domain structures and significant
hysteresis make such films a convenient paradigm for analyzing the
microstructural origin of magnetic hysteresis \cite{Ber}.

Most current modeling of soft thin films is based on direct micromagnetic
simulation \cite{micromagnetists}. This is demanding due to the 
long-range nature of dipolar interactions, and the necessity of 
resolving several small length scales simultaneously. 
Numerical simulation is surely the right tool for the quantitative study
of hysteresis and dynamic switching \cite{reversal}. However it is
natural to seek a more analytical understanding of the equilibrium
configurations. The origin of domain patterns is intuitively clear:
they arise through a competition between the magnetostatic effects
(which favor pole-free in-plane magnetization) and the applied field
(which tends to align the magnetization). A 2D model based on this
intuition was developed by van den Berg \cite{VdB} in the absence of an
applied field, and extended by Bryant and Suhl \cite{BS} to the case of a 
sufficiently weak in-plane applied field. In van den Berg's model (VDBM)
magnetic domain patterns are represented using 2D, unit-length, divergence-free
vector fields, determined using the method of characteristics; the
caustics where characteristics meet are domain walls. In Bryant and Suhl's
model (BSM), the presence of a weak applied field is accounted for through
an electrostatic analogy: the ``charges'' associated with the magnetic
domain pattern should be such as to expel the applied field from the interior
of the sample, as occurs in an electrical conductor. The domain patterns
predicted by BSM have been observed experimentally \cite{RH}. 
The electrostatic analogy is 
restricted, however, to sufficiently small applied fields: since
the magnetization vector has a constrained magnitude, the field generated
by its divergence cannot be arbitrarily large. Therefore the BSM breaks
down at a critical field strength beyond which the external
field penetrates the sample.

This Letter extends and clarifies the models of van den Berg and
of Bryant and Suhl.
Our extension is two-fold: we permit large applied fields
which penetrate the sample, and we replace the method of characteristics
with a robust numerical scheme. Our clarification is also two-fold:
we identify the regime in which these 2D models are valid, and explain
their relation to classical micromagnetics. To assess the extended model,
we compare its predictions to experiments on Permalloy thin film
elements with square cross-section. The agreement between theory and
experiment is remarkable, even in the field penetration regime.
At the heart of our approach is an asymptotic analysis of the micromagnetic
energy in the thin-film limit. The lowest-order terms lead to constraints such
as $m_3=0$, while the second-order term sets the charge density. Wall 
energies and anisotropy contribute only at higher order. The higher-order
terms are not irrelevant: they are the source of magnetic hysteresis.
Our analysis indicates, however, that certain quantities should have
little or no hysteresis -- namely the charge density, the region of field
penetration, and the magnetization in the penetrated region.

The free-energy functional of micromagnetics in 
units of $J_{s}^{2} L^{3} / 2 \mu_{0}$ is 
\begin{eqnarray}\label{3d-energy}
E_{d}(m) &=& (\kappa d)^{2} \int_{\Omega_{d}} |\nabla m |^{2} dx +
Q  \int_{\Omega_{d}} \varphi (m)dx \nonumber \\
& +  &\int_{\R^{3}} |h_{d}|^{2}dx-
2 \int_{\Omega_{d}} h^{\prime}_{e} \cdot m dx\,. 
\end{eqnarray}
Here $m$ is the magnetization
(in units of the saturation magnetization $J_s$),
a unit vector field defined
on the film $\Omega_{d}$ with cross section $\w$ and thickness
$d$, where all lengths are measured in units of a typical lateral dimension $L$
(the diameter for $\w$ a circle, the edge-length for $\w$ a square).
Moreover, $\kappa$ is the ratio between Bloch line width $D_{BL}$ and the film thickness,
where $D_{BL}=(2\mu_{0}A/J_{s}^{2})^{\frac{1}{2}}$, with $A$ the exchange constant,
measures the strength of the exchange energy relative to that
of  dipolar interactions; $Q$ is the quality
factor measuring the relative strength of the magnetic anisotropy $\varphi$;
$h_{d}$ is the  stray field in units of $J_{s}/\mu_{0}$,
whose norm squared gives the magnetostatic energy
density; $h^{\prime}_{e}$ is the applied field in units of $J_{s}/\mu_{0}$,
which we assume to be uniform and parallel to the film's cross section.
In what follows, a prime will always denote a two-dimensional field or operator.

For $d\ll 1$ a hierarchical structure emerges
in the energy landscape of (\ref{3d-energy}), see Table \ref{scalings}.
Variations of $m$ of order $1$ along the thickness direction $x_3$
give rise to an exchange energy per unit area  (of the cross section) of
order $\kappa^2 d$. An out--of--plane component $m_3$ of order one
determines a magnetostatic contribution per unit area of
order $d$. The component of the in-plane magnetization $m^\prime$ orthogonal
to the lateral boundary $\partial\omega$ of the film's cross section
$\omega$ leads to a magnetostatic contribution of order $d^2 \ln \frac{1}{d}$
per unit length. The same mechanism penalizes jumps 
$\left[ m^\prime\cdot\nu^\prime \right]$
of the normal component of the magnetization across a line of
discontinuity of $m^\prime$ with normal $\nu^\prime$. These lines
of discontinuity arise by approximating domain walls as sharp interfaces.
At order $d^2$ we find the magnetostatic energy per unit area due to surface
``charges'' proportional to the in-plane divergence $\div^{\prime}m^{\prime}$.
Finally, the energy per unit length of a N\'eel or asymmetric
Bloch wall and the energy of a single vortex are 
indicated in the table.
In the regime
\begin{equation}\label{materparam}
H'_e=\frac{h'_e}{d}\sim1,\quad\frac{Q}{d}\ll1,\quad
d\ll\kappa^2\ll\frac{1}{d\ln \left( \frac{1}{d} \right)} \,,
\end{equation}
the highest-order terms penalizing $m_3$, $\frac{\partial m}{\partial x_3}$ and
$[m'\cdot\nu']$ become hard constraints, while the energetic cost
of anisotropy, of the wall type of minimal energy \cite{walltypes}, 
and of vortices become
higher-order terms. The energy is thus determined, at principal order,
by the competition between the aligning effect of $H^{\prime}_{e}$
and the demagnetizing effects due to $\div^{\prime}m^{\prime}$.

\begin{table}
\caption{Scaling of various energy sources}
\begin{center}
\begin{tabular}{cc}
$\frac{\partial m}{\partial x_3}$ & $\kappa^{2}d$ \\
$m_3$ & $d$ \\
$\left[m'\cdot\nu'\right]$ & $\ln(\frac{1}{d})\,d^2$ \\
$\div^{\prime}m^{\prime}$ & $d^2$ \\
external field energy &  $h'_e\,d$ \\
anisotropy energy & $Q\,d$ \\
asymmetric Bloch wall & $\kappa^2\,d^2$ \\
N\'eel wall &
$(\ln(\frac{1}{\kappa^2\,d}))^{-1}\,d^2$ \\
vortex & $\ln(\frac{1}{\kappa\,d})\,\kappa^2\,d^3$
\end{tabular}
\end{center}
\label{scalings}
\end{table}

In view of this separation of energy scales in the regime
(\ref{materparam}), we propose the following reduced theory.
We call an in--plane vector field $m'(x')$ on $\omega$ ``regular'' if
it satisfies $\left[m^{\prime}\cdot\nu^{\prime}\right]=0$ across
all possible discontinuity lines and at $\partial\w$. 
Our reduced theory states that the magnetization $m'(x')$ minimizes
\begin{equation}\label{2d-energy}
E(m^{\prime})=\int_{\R^3}|H_{d}|^2\,dx
-2\int_{\omega}H^{\prime}_{e}\cdot m^{\prime}\,dx^{\prime},
\end{equation}
where $H_{d}(x)=-\nabla U$ is determined by
\begin{eqnarray*}
\nabla^{2} U&=&0\;\;\mbox{in}\;\R^3\;\mbox{outside of}\;\w,\\
\left[ \frac{\partial U}{\partial x_{3}} \right]&=&
{\rm div}^{\prime} m^{\prime}\;\;\mbox{on}\;\omega\,,
\end{eqnarray*}
among all regular in-plane vector fields $m^{\prime}$ of unit length
\begin{equation}\label{unit}
|m^{\prime}|=1\;\;\mbox{in}\;\omega.
\end{equation}
Our formula for the induced field $H_d$ is naturally consistent with
that commonly used for 2D micromagnetic simulations \cite{2Dstray}.

We now make two crucial observations. The first is that
the functional $E$ depends on $m'$ only via the surface charge
$\sigma = - {\rm div}'m'$, and it is strictly convex in $\sigma$.
Indeed, $\int_{\R^3}|H_d|^2\,dx$ is a quadratic functional of 
$\sigma$ and an integration by parts shows that
$\int_\omega H_e'\cdot m'\,dx'$ is a linear functional of $\sigma$.
The second observation is that the set of regular in--plane vector
fields of unit length and with given surface charge 
is large in the following sense: For any regular 
$m'_0$ of at most unit length, that is
\begin{equation}\label{subunit}
|m'_0|\le1\;\;\mbox{in}\;\omega,
\end{equation}
there exist many regular $m'$ of unit length with the same
surface charge: ${\rm div}'m'={\rm div}'m'_0$. Indeed, we may write
$m'=\nabla^\perp\psi+m'_0$ where
$\nabla^\perp\psi=(-\partial\psi/\partial x_{2},\partial\psi/\partial x_{1})$
and the continuous function $\psi(x')$ on $\omega$ solves the boundary
value problem
\begin{eqnarray}\label{HJ1}
|\nabla^\perp\psi+m'_0|&=&1\;\;\mbox{in}\;\omega,\\
\psi&=&0\;\;\mbox{on}\;\partial\omega.\label{HJ2}
\end{eqnarray}
Condition (\ref{subunit}) ensures the solvability of this boundary 
value problem. One can generate many solutions by imposing
the additional condition $\psi=0$ on an arbitrary curve
contained in $\omega$.

These observations have two important consequences.
First, the minimizer of the reduced energy $E$ is not uniquely
determined. Indeed, according to our first observation, $E$ depends
only on the surface charge, and according to our second observation, a 
regular in--plane vector field of unit length is not uniquely determined
by its surface charge.

The second consequence is that the surface charge and thus the
stray field are
uniquely determined. Indeed, according to our first observation, $E$
is a strictly convex function of the surface charge, and according to
our second observation, the set of surface charges which can be generated by
regular in--plane vector fields of unit length is convex. (This is true
despite the fact that the set of regular in--plane vector fields with
unit length is not convex.)

Any minimizer $m^{\prime}$ of (\ref{2d-energy},\ref{unit}) 
satisfies the Euler-Lagrange equation
\begin{equation}\label{el}
H^{\prime}_{d}+H^{\prime}_{e}=
\lambda\,m^{\prime}\;\;\mbox{in}\;\omega,
\end{equation}
where $\lambda(x^{\prime})$ is the Lagrange multiplier associated 
with the pointwise constraint (\ref{unit}). Since $H_d$ is uniquely
determined, the region $\{H^{\prime}_{d}+H^{\prime}_{e}\not=0\}$
of $\omega$
where the external field is not expelled from the sample is uniquely
determined. Within this penetrated region, $m'$ is also uniquely
determined in view of (\ref{el}).

There is a finite critical field strength $H_{crit}$,
in the following sense: when the applied field is subcritical $\lambda \equiv 0$
and the field is completely expelled from the sample, whereas when it is
supercritical $\lambda$ is nonzero somewhere and the field penetrates
in that  part of the sample.
The critical field strength depends on the geometry of $\omega$ --- for 
a circular disk of diameter $1$, its value is $1$. 
Further analysis indicates that there can be no walls (discontinuity lines of
$m'$) in the penetrated region. Moreover the penetrated region must meet
the boundary of $\omega$.

To derive quantitative predictions from our reduced model
(\ref{2d-energy},\ref{unit}), we proceed in two steps. 
The first step minimizes (\ref{2d-energy}) among all regular
in--plane vector fields $m'_0$ of length {\em less than or equal to} $1$.
Recall that replacing (\ref{unit}) by (\ref{subunit}) does not change the
minimum energy; therefore the $m'_0$ obtained this way has the
correct reduced energy, though it typically violates (\ref{unit}). The
second step postprocesses $m'_0$ by solving (\ref{HJ1},\ref{HJ2}) to obtain
another minimizer $m'$ of unit length. This $m'$ is the desired
magnetization.

The first step is a convex (though degenerate) variational problem. We
solve it using an interior point method \cite{vanderbei}:
the convex constraint is enforced by adding to the physical energy $E$
a small multiple $t$ of a self--concordant barrier $B$.
The unique stationary point of the strictly convex $E+tB$ is computed by
Newton's method; it serves as an initial guess for the
minimizer of $E+t'B$, where $t'<t$. The parameter $t$ is slowly
decreased by multiplicative increments. Within Newton's method, 
the Hessian of $E+tB$ is inverted by a preconditioned conjugate
gradient method. The magnetostatic part of the Hessian is evaluated
with the help of FFT. This is a robust procedure.

For the second step, we recall that the solution of
(\ref{HJ1},\ref{HJ2}) is not unique. However there is a special solution
$\psi$, known as the ``viscosity solution'', which has
special mathematical properties \cite{evans}. It is robust and can be
computed efficiently using the ``level set method'' \cite{S}. This is what
we compute.

Our numerical scheme selects --- automatically and robustly --- one
of the many minimizers $m'$. The selection principle implicit in this
scheme is the same as the one proposed by Bryant and Suhl. 
It appears to pick a minimizer with as few walls as possible. Thus it 
is not unlike the more physical selection mechanism of minimizing wall energy,
represented as a higher-order correction to (\ref{2d-energy}) \cite{AG}.

Figure \ref{fig:num} shows the predictions of our numerical scheme 
for a square film of edge-length one, subject to a monotonically
increasing field applied along the diagonal. To check our predictions,
we have observed the response of two ac-demagnetized Permalloy
(Ni$_{81}$Fe$_{19}$, $J_s=1.0$ T) square samples of edge 
lengths $L=$ 30 and 60 $\mu$m and thicknesses $D=$ 40 and 230 nm,
respectively, in a digitally enhanced Kerr microscope.
The observed domain 
patterns are given in Figures \ref{fig:expthick}, \ref{fig:expthin} where
the field intensity $h_{e}$, measured in Tesla, is scaled according to
\begin{equation}
H =  \frac{L}{D} \frac{h_{e}}{J_{s}}.
\end{equation}
\begin{figure}[t]
\begin{figure}[t]
  \begin{center}
    \leavevmode
    \psfig{figure=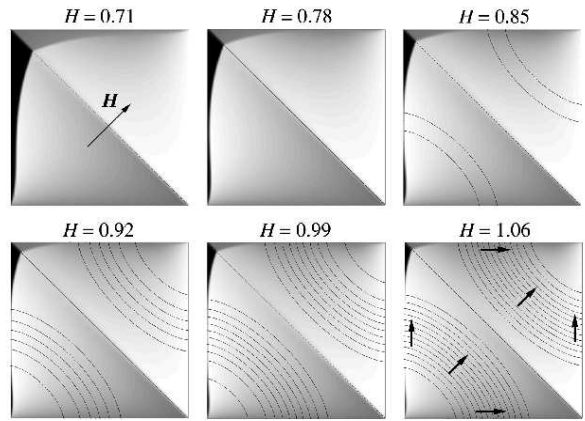,width=0.93\columnwidth}
  \end{center}
  \caption{The transition between expulsion and penetration regimes.
  In the last picture, arrows indicate the magnetization direction.}
  \label{fig:crit}
  \label{fig4}
\end{figure}

  \begin{center}
     \leavevmode
     \psfig{figure=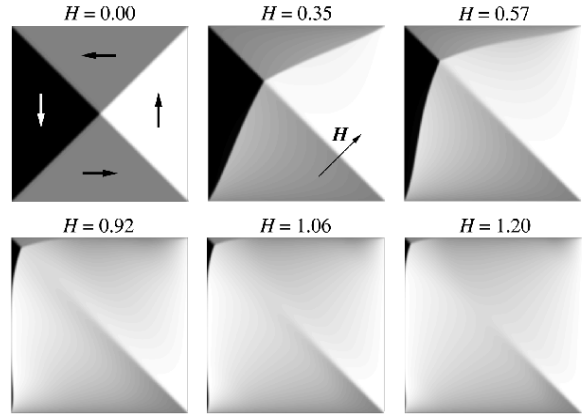,width=0.93\columnwidth}
  \end{center}
  \caption{Predictions of the theory: gray-scale plots of the vertical
component of magnetization.}
  \label{fig:num}
  \label{fig1}
\end{figure}
\begin{figure}[t]
  \begin{center}
    \leavevmode
    \psfig{figure=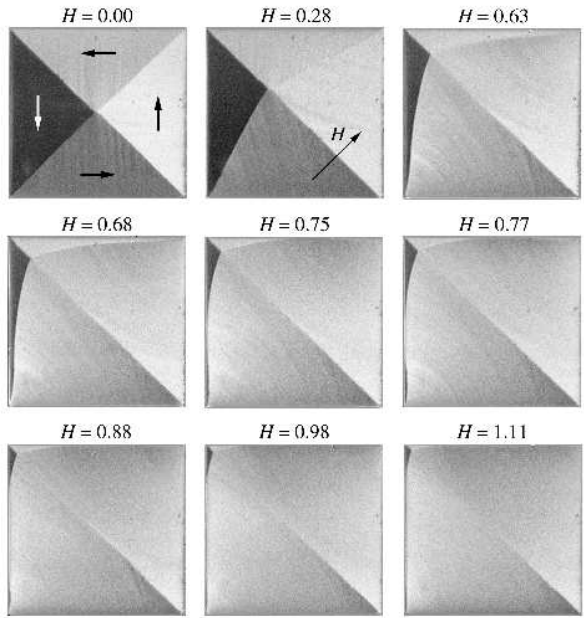,width=0.93\columnwidth}
  \end{center}
  \caption{Permalloy films: $L=60$ $\mu$m, $D=230$ nm.}
  \label{fig:expthick}
  \label{fig2}
\end{figure}
\begin{figure}[t]
  \begin{center}
    \leavevmode
    \psfig{figure=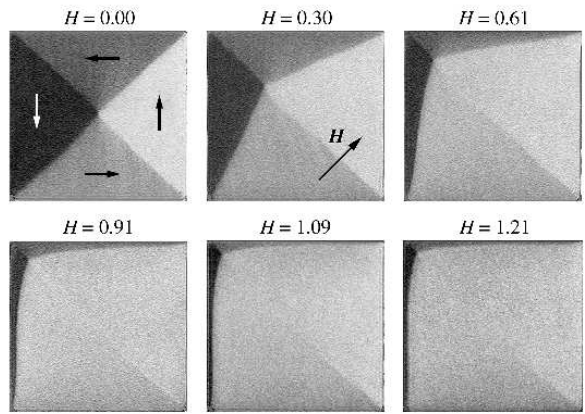,width=0.93\columnwidth}
  \end{center}
  \caption{Permalloy films: $L=30$ $\mu$m, $D=40$ nm.}
  \label{fig:expthin}
  \label{fig3}
\end{figure}
Figure \ref{fig:crit} examines more closely the predictions of
our theory for  $|H'_e|$ close to $H_{crit}$.
We have superimposed on each gray-scale plot the level curves of the
potential $v$ of the penetrated field, defined by
$$
-\nabla v = H^{\prime}_{d}+H^{\prime}_{e}\,.
$$
Regions where the field lines concentrate are regions where $\nabla v \neq 0$,
i.e., where the external field has penetrated the sample.
Within them, (\ref{el}) implies that
$m^{\prime}$ is parallel to $\nabla v$ .
Our theory predicts that the penetrated region must
meet the boundary of the sample and that 
$m'$ can have no walls in the penetrated region.
The pictures confirm this, and show quite clearly that two apparently independent
phenomena -- the expulsion of the domain walls from the interior of the sample
and the penetration of the external field --
are in fact two manifestations of the same event.

In summary, our model describes the response of a soft ferromagnetic
thin film to an applied magnetic field. It determines the micromagnetic
energy to principal order, and certain associated physical quantities
that should have little or no hysteresis --- the charge density, the
region of field penetration, and the magnetization in the penetrated
region. In addition our approach provides a specific magnetization
pattern which is consistent with experimental observation and may well
be the ground state. Of course, the magnetization of a soft thin film
is not uniquely determined by the applied field: the multiplicity of
metastable states is a primary source of hysteresis. Our approach
does not provide a model for hysteresis or a classification of 
stable structures --- this would seem to require analysis of 
higher-order terms in the micromagnetic energy.

We thank S. Conti for valuable discussions.


\begin{references}
\bibitem{walltypes}
Note that 
(\ref{materparam}) probes a range of film thicknesses over
which different wall types are to be expected, see ref. \cite{HS}. 
Typical values
for Permalloy are $Q=2.5\times 10^{-4}$ and $D_{BL}=5$nm. For a
circular element with diameter $1$ $\mu$m and thickness $10$ nm
we have $d=0.01$ and $\kappa =0.5$.
\end{references}
\end{document}